**Fisheries Management in Congested Waters: A Game-Theoretic Assessment of the East China Sea**

Michael Perry, mperry20@gmu.edu
George Mason University, Department of Systems Engineering and Operations Research, Fairfax, VA, USA

**1. Introduction**

The East China Sea (ECS) is a congested maritime environment in the sense that fishermen from multiple nations can access these waters with essentially equivalent operating costs. A journey from China to Japan through the ECS, for example, can span less than 350 nautical miles, while one from China to South Korea can span only 200 nautical miles. As a point of fact, fishermen from each of these three nations have historically fished throughout the ECS under less restrictive fisheries management regimes. Other congested maritime environments can of course be identified, such as the South China Sea (SCS) where disputes exist between all major SCS nations, but a distinguishing feature of the ECS is that legal agreements have been made on paper for who's allowed to fish where (Rosenberg 2005).

Despite such agreements between each China, South Korea, and Japan, illegal encroachments persist in high volumes (Hsiao 2020). Writing this off as the behavior of self-interested criminals is unsatisfactory, as state-led monitoring, controls, and surveillance (MCS) of fisheries present cost-effective ways to deter illegal fishing, and yet MCS has been rated poorly among ECS nations by the United Nations Food and Agricultural Organization (FAO) (Petrossian 2019). If these illegal encroachments were a true concern of the state, one would see more investment in MCS. This paper provides a game-theoretic explanation for how rational state policy can allow illegal encroachments to persist.

The paper analyzes a congested maritime environment using a game with three players: Japan, South Korea, and China. A key modeling framework is that despite legally defined fishing waters, nations can tacitly allow illegal fishing by issuing excessive quotas in their own waters while letting economic principles determine whether each fisherman will fish in Japanese waters, South Korean waters, Chinese waters, or none at all. Coupled with a standard fisheries model, this economic determination is influenced by asymmetries in costs. While the proximities of nations are assumed to render operating costs equivalent, asymmetries exist in the ECS on at least two accounts. Opportunity costs are the most obvious asymmetry, as income opportunities in China differ greatly from South Korea and Japan. The other source of asymmetry addressed is that arising from the deterrent effect of patrol crafts. Not only can nations have differences in the quantity and quality of patrols, which impose costs on illegal fishing, but a nation's fishermen also differ in their ability to resist patrols. The latter point has garnered significant attention in recent years with the rise of so-called maritime militias; that is, fishermen armed with martial assets (often with state sponsorship) such as water cannons and small arms that embolden them to fish in illegal waters. As will be shown, the low-cost player can extract rent from the higher-cost player's waters by issuing excessive quotas; one strategy high-cost players can use to combat this is to issuing excessive quotas themselves, reducing the attractiveness of their waters to illegal fishermen and creating a vicious circle of overfishing.

As already alluded, fisheries management entails more than issuing the "right" amount of quotas. MCS measures are also essential to ensure fishermen are fishing only where allowed, catching the allowable species in the allowed quantities, and so on. For the purposes of this paper, MCS refers to such deterrent practices other than maritime patrols and includes measures such as port inspections and vessel monitoring systems.  It does not include sound practices for issuing quotas, which is treated as a separate decision variable.  The poorly-rated MCS observed in the ECS can also be explained through strategic interaction in a congested environment.  It's seen that even if MCS is a cost-effective deterrent of illegal fishing in the single-state context, once additional players are introduced the incentive to invest in MCS vanishes.

A critical component in the analysis is that each player can behave against the norms of agreed upon fishing rights while maintaining plausible deniability.  For example, a country, say China, cannot explicitly authorize fishermen to fish in South Korean waters, as if this were discovered the political consequences would be unacceptable.  This assertion is consistent with observed behavior as China has historically punished its fishermen caught in South Korean waters (Zhang 2016).  Likewise, if it were discovered China were knowingly letting fish caught in South Korean waters in through its ports, the consequences would be unacceptable.  It might be asked, then, what is to be done, if an agreement is already on paper but countries nonetheless tacitly violate it?  To address this, the paper assesses what enforceable policy on fishing quotas can be implemented to improve each player's utility.  Such a policy involves an observable agreement on legal fishing levels, including legally allowed fishing by one player in another's waters, as well as mutually observable MCS regimes.  Unsurprisingly, the low-cost player, either by virtue of low opportunity costs or strong patrols and maritime militias, has leverage in negotiating such an enforceable policy.  However, a Nash bargaining problem is analyzed to show that all players benefit substantially from negotiation when benchmarked against the noncooperative alternative.

The remainder of the paper is structured as follows. Section 2 reviews the literature on fisheries management in the ECS as it pertains to illegal fishing, game theoretic models of fisheries, and the deterrent effects of maritime law enforcement on illegal fishing. Section 3 describes a model for the ECS as a multi-player game where each player owns a specified number of fisheries; this includes a subgame to determine where fishermen with quotas will fish. Analytical results on the existence and uniqueness of subgame equilibria are derived, as is a result showing the optimal response of all players is to use no MCS. Section 4 presents examples and comparative statics. Section 5 concludes and comments on future work. An appendix is provided with mathematical proofs for all theorems.

## 2. Literature Review

Bilateral agreements delineating fishing rights in the ECS began to take form in the late 1990s, delineating regions that would fall under the jurisdiction of particular states as well as jointly managed areas. These agreements have also included provisions for legal fishing by one country in the waters of another (Rosenberg 2005; Hsiao 2020). However, these agreements are not fully cooperative in the sense that they haven't set limits on domestic fishing; that is, a bilateral agreement between China and South Korea may specify legal Chinese fishing levels in South Korean waters, but not legal Chinese levels in Chinese waters. This oversight has led to overfishing, and is perhaps the cause of current cutbacks on legally allowed foreign fishing (*Yonhap News* 2020). Despite these agreements, illegal encroachments persist in large numbers. An estimated 29,600 instances of Chinese fishing vessels entering South Korean waters occurred in the second half of 2014 alone. Further, and in support of the assertion states may only tacitly encourage illegal fishing but not outright allow it, Chinese officials have taken punitive action against its illegal fishermen when detained by other countries (Zhang 2016). Analyzing these disputes has been complicated by the rise of maritime militias, the most well-studied of which is the Chinese Maritime Militia. These militias are fishermen who've been equipped with martial assets such as small arms and water cannons, generally through state-sponsorship, to oppose other nation's fishermen and patrol craft (Erickson and Kennedy 2016; Zhang and Bateman 2017; Perry 2020). In the most extreme cases, interactions between fishermen and maritime law enforcement has led to the loss of life on both sides (China Power Team 2020; Park 2020).

A key tool for addressing illegal encroachments, and illegal fishing in general, is monitoring, controls, and surveillance (MCS), which includes measures such as inspections at ports of entry, onboard observers to monitor fishing activities, and satellite systems to track vessels (Pitcher, Kalikoski, and Pramod 2006). The definition of MCS also often includes the issuance of quotas, but quotas issuance is treated separately in this paper. Empirical studies have shown MCS to be an effective deterrent to illegal fishing, and while costs can vary widely irrespective of MCS quality its generally viewed as a cost-effective tool (Petrossian 2015; Mangin et al. 2018). Fisheries management experts have, nonetheless, rated MCS practices among ECS nations as below average (Petrossian 2019). Interestingly, studies have also found more generally that fisheries management tends to be weaker when nearby states are also not adhering to best practices, lending credence to the modeled results in this paper for congested maritime environments (Borsky and Raschky 2011).

From a modeling perspective fisheries disputes have often been studied through game theory. In the well-studied problem of two or more players allocating quotas in the same fishery, models have shown overfishing will occur both when considering discounted payoffs and not. When accounting for asymmetric costs a player's utilities are seen to increase as another's costs increase (Grønbæk et al. 2020). While related to the findings of this paper, the distinction is the model presented here considers multiple fisheries with legal definitions of who's allowed to fish in each, introducing a criminological element into the analysis. Models have accounted for additional complicating factors such as multiple interacting species (Fischer and Mirman 1996), coalition forming in many player games (Long and Flaaten 2011), and uncertainty in stock levels and growth (Miller and Nkuiya 2016), to name a few. Models have also been used to optimize patrol strategies to combat illegal fishing. These models discretize a nation's waters and consider two players, the state and illegal fishermen, who must determine how to deploy patrols and where to attempt illegal fishing, respectively. Factors such as heterogeneous illegal fishermen types, bounded rationality, and repeated play have all been incorporated (Fang, Stone, and Tambe 2015; Brown, Haskell, and Tambe 2014). What has not been considered is situations with multiple states competing over resources utilizing either state-sponsored or tacitly encouraged illegal fishing.

Regarding the economics of illegal fishing, researches have assessed the annual dollar amount of illegally caught fish to be between $10-$23.5 billion (Petrossian 2019). In addition to MCS, empirical studies have found the number of patrol craft per 100,000 square kilometers of water to be a significant explainer of illegal fishing (Petrossian 2015). Deterrence has also been analyzed at the agent-level, where studies have shown fishermen do indeed rationalize their illegal fishing by pointing out the risks and costs associated with being caught don't outweigh the benefits (King and Sutinen 2010; Kuperan and Sutinen 1998; Nielsen and Mathiesen 2003). Lastly, note a gap in the empirical literature exists in understanding the impact of maritime militias on the willingness to fish illegally; while maritime militias have been studied extensively, no empirical evidence exists to quantify their impact.

## 3. The Congested Environment Fishing Game

The scenario is modeled as a multi-player game where each player has ownership of one or many fisheries. Each fishery is modeled using the Gordon-Schaeffer model with a logistic growth function and constant-effort harvesting rate (Clark 2006; Schaefer 1954; Gordon 1954). The model for fishery $j \in \mathbb{Z}^+$ of player $i$, which will be referred throughout this paper as fishery $ij$, is:

$$\frac{dx_{ij}(t)}{dt} = r_{ij} x_{ij}(t) \left(1 - \frac{x_{ij}}{Z_{ij}}\right) - x_{ij}(t) q_{ij} \sum_{kl} F_{ij,kl}, \tag{1}$$

where $x_{ij}(t)$ is the biomass in the fishery at time $t$, $Z_{ij} \in \mathbb{R}^+$ is the carrying capacity, $r_{ij} \in \mathbb{R}^+$ the natural growth rate, and $q_{ij} \in \mathbb{R}^+$ the catchability coefficient. $F_{ij,kl} \in \mathbb{R}^+$ is the level of fishing effort in the fishery by fishermen who are authorized to fish in fishery $kl$ (the $l^{th}$ fishery of player $k$). In other words, the variable $F_{ij,kl}$ for $i \neq k$ or $j \neq l$ represents illegal fishing, whereas $F_{ij,ij}$ represents legal fishing. No migration between fisheries is assumed. In this paper long-term biomass in each fishery is used to evaluate utilities. Given all values of $F_{ij,kl}$, the steady-state solution for biomass, $x_{ij}$, in the differential equation (1) is:

$$x_{ij} = Z_{ij} \left[1 - \frac{q_{ij} \sum_{kl} F_{ij,kl}}{r_{ij}}\right]. \tag{2}$$

To reiterate the key modeling framework, each player's national strategy merely determines how many fishing quotas are allocated in their respective fisheries, denoted $F_{kl}$ for each fishery $kl$. It's assumed that without any explicit instruction from the state, individual fishermen with quotas decide where to fish based on economic principles. The variables $F_{ij,kl}$ will be determined via a subgame to be detailed momentarily.

An individual fishermen's revenue is a function of biomass. Denoting the price of fish farmed from fishery $ij$ as $p_{ij} \in \mathbb{R}^+$, the revenue earned by a fisherman in this fishery is $p_{ij} q_{ij} x_{ij}$. That fishermen's rent is determined by subtracting out costs, which obviously includes operational costs, but also asymmetric opportunity costs and costs imposed through the efficacy of maritime patrols to deter illegal fishing. It's assumed all fisheries are in close enough proximity such that operating costs are the same in each and for each player. In sum, the costs for a fisherman to fish in $ij$ when he's legally authorized to fish in $kl$ is defined as follows:

$$c_{ij,kl} = \begin{cases} c_k, & \text{if } i = k \text{ and } j = l \\ c_k + \beta_k P_i + \beta_m m_k, & \text{otherwise} \end{cases}, \quad (3)$$

where $c_k \in \mathbb{R}^+$ is the combined operating and opportunity costs for fishermen from country $k$, $P_i \in \mathbb{R}^+$ is the number of patrol craft employed per 100,000 square kilometers by country $i$, and $m_k \in \mathbb{R}^+$ is the level of MCS used in country $k$ (consistent with studies such as Petrossian (2015), $m_k$ is measured as an expert's assessment of MCS quality). The coefficient $\beta_m \in \mathbb{R}^+$ represents the deterrent effect of MCS on illegal fishing and is assumed to be the same for all players. $\beta_k \in \mathbb{R}^+$, in contrast, is the deterrent effect of patrols and is allowed to vary between players. This reflects differing levels of maritime militias among ECS nations. Considering these costs, the rent earned by a fisherman is $\pi_{ij,kl} = p_{ij} q_{ij} x_{ij} - c_{ij,kl}$.

Total utility realized by player $k$, $u_k$, is the sum total of rents collected by their nation's fishermen, minus expenditures on MCS. MCS costs are modeled as a function of the level of MCS used; the flexible form $a_1 m_k^{a_2}$ is used, where $a_1, a_2 \in \mathbb{R}^+$ are constants. The level of patrols is treated as a given constant rather than a decision variable, as in practice many other considerations affect investment in patrol craft; for instance, port security, detection of human trafficking, and search and rescue operations, to name a few applications. The utility function for player $k$ is therefore:

$$u_k = \sum_{ij} \sum_{kl} \pi_{ij,kl} - a_1 m_k^{a_2} \qquad (4)$$

By solving a subgame for the values of $F_{ij,kl}$, this utility function collapses into an expression of the player's overall decision variables ($F_{kl}$ and $m_k$).

*3.1. Subgame for Levels of Fishing in Each Fishery*

A state-issued fishing quota gives fishermen the legal right to bring fish into the country of issue. States cannot explicitly direct fishermen to fish illegally; the decision on where to fish in a congested environment is therefore left to the fishermen, a choice which will be made based on economic principles. Elementally, the following must hold for a fishermen's choice to be economically rational: (i) a fisherman will use a quota iff he can earn nonnegative rent; (ii) to fish in fishery $ij$, a fisherman must be earning at least as much as he could in any alternative fishery, $mn$, as otherwise he would switch to the former. These subgame principles can be modeled by the below binary program, where the legally authorized quotas in each fishery, $F_{kl}$, and the levels of MCS used, $m_k$, are given first-stage decision variables. The binary program finds values of the subgame variables (SGVs) $F_{ij,kl}$ such that points (i) and (ii) are satisfied for all fishermen.

$$\min_{w,\, F_{ij,kl},\, y_{ij,kl},\, y_{kl}} w \qquad (5)$$

s.t.

*Noninformative objective function* (5a)
$$w \geq 0$$

*Constraints on total fishing* (5b)
$$\sum_{ij} F_{ij,kl} < F_{kl} \; \forall \; kl$$

*Fishing only occurs where profitable, and is at least as profitable as any alternative fishery* (5c)
$$F_{ij,kl} \leq M y_{ij,kl} \; \forall \; ij,\, kl$$
$$F_{ij,kl} \geq .0001 y_{ij,kl} \; \forall \; ij,\, kl$$
$$\pi_{ij,kl} \geq -M(1 - y_{ij,kl}) \; \forall \; ij,\, kl$$
$$\pi_{ij,kl} \geq \pi_{mn,kl} - M(1 - y_{ij,kl}) \; \forall \; ij,\, kl,\, mn$$

*No profitable quotas are left unused* (5d)
$$\sum_{ij} F_{ij,kl} + M y_{kl} \geq F_{kl} \; \forall \; kl$$
$$\sum_{ij} F_{ij,kl} + .0001 y_{kl} \leq F_{kl} \; \forall \; kl$$
$$\pi_{ij,kl} \leq M(1 - y_{kl}) \; \forall \; ij,\, kl$$

*Nonnegativity and binary constraints* (5e)
$$F_{ij,kl} \geq 0 \; \forall \; ij,\, kl$$
$$y_{ij,kl},\, y_{kl} \in \{0,1\}.$$

The parameter $M$ is an arbitrarily large constant. In constraints (5c), notice that if $F_{ij,kl} > 0$, the only feasible value for $y_{ij,kl}$ is 1, which in turn enforces $\pi_{ij,kl} \geq 0$ and $\pi_{ij,kl} \geq \pi_{i'j',kl} \; \forall \; k'l'$. In contrast, $F_{ij,kl} = 0$ enforces $y_{ij,kl} = 0$, which in turn places no meaningful constraint on $\pi_{ij,kl}$ (since $M$ is arbitrarily large). In constraints (5d), an instance of unused quotas, $\sum_{ij} F_{ij,kl} < F_{kl}$, enforces $y_{kl} = 1$ and $\pi_{ij,kl} \leq 0$, so that remaining opportunities for positive rent exist. If all quotas are used, $\sum_{ij} F_{ij,kl} = F_{kl}$, then $y_{kl} = 0$ is the only feasible choice. The objective function of (5) is irrelevant, as the purpose of the subgame is to find feasible values of the SGVs. It can be proven that a subgame equilibrium (SGE) always exists (see Theorem 1), and while they're not unique, all SGE yield identical utilities for the players, making the specific SGE used to analyze the overall game irrelevant (see Theorem 2).

**Theorem 1. Existence of a subgame equilibrium.**

Any instantiation of the game's parameters and choice of the overall game's decision variables yields a subgame equilibrium.

*Proof.* See Appendix A.

**Theorem 2. Utilities are uniquely defined by decision variables.**

Any instantiation of the game's parameters and choice of the overall game's decision variables leads to uniquely defined utilities for the players. This is in spite of the fact that multiple subgame equilibria may exist.

*Proof.* See Appendix A.

*3.2. Optimality of $m_k = 0$*

Having stated the subgame, to find an equilibrium for the overall game one needs to find values of $F_{kl}$ $\forall$ $kl$ and $m_k$ $\forall$ $k$ such that no player can increase their utility from a unilateral move. Various methods are used in Section 4 to find such strategy profiles (depending on whether the players each own multiple fisheries), but finding them is made easier by the following result, which states it's never optimal for states to employ MCS in a noncooperative game where each player owns only one fishery. Note that, experimentally, this result appears to be true in the case where each player owns multiple fisheries as well, as seen in Section 4.

**Theorem 3. Non-use of MCS.**

Assume Blue and Red each own one fishery. Consider Player $k$'s decision. For given strategies of all other player, responding with $m_k > 0$ can yield at most equivalent utility as $m_k = 0$. If costs of MCS are nonzero, then $m_k > 0$ yields strictly less utility than $m_k = 0$.

*Proof.* See Appendix A.

It's worth recalling the definition of MCS used in this paper: measures taken to ensure fishing is occurring where allowed, other than patrols. In other words, $m_k = 0$ still allows for some degree of strong fisheries management, such as strictly managed quota systems, restrictions on access to fishing gear, and patrols. Not using MCS amounts to giving fishermen with a legal quota a free pass to bring fish in through ports, irrespective of where it was caught.

## 4. Examples

### 4.1. Example 1: One Fishery per Player

This initial example analyzes a three-player game between Japan, South Korea, and China where each owns only one fishery. This most basic instantiation of the model is sufficient to understand why nations overfish, and why illegal encroachments persist, in a congested environment such as the ECS. Since each player owns only one fishery the prior notation of "$ij$" and "$kl$" will be condensed to simply "$i$" and "$k$", where $i, k \in \{J, S, C\}$ to indicate Japan, South Korea, and China, using the obvious abbreviations. The below parameters are used, reflecting higher opportunity costs in the relatively wealthy nations of Japan and South Korea, and the relatively high levels of investment in patrols and maritime militias by China (Erickson 2018). Further justification for parameter choices is provided in Appendix B. All biological parameters are measured in millions of metric tons of fish, and all monetary parameters are in billions of USD:

$$Z := Z_J = Z_S = Z_C = 2$$

$$r := r_J = r_S = r_C = .4$$

$$q := q_J = q_S = q_C = .0002$$

$$p := p_J = p_S = p_C = 3$$

$$c_J = 250 \times 10^{-6}, c_S = 200 \times 10^{-6}, c_C = 120 \times 10^{-6}$$

$$P_J = 20, P_S = 30, P_C = 50$$

$$\beta_J = 6 \times 10^{-7}, \beta_S = 6 \times 10^{-7}, \beta_C = 4 \times 10^{-7}$$

$$\beta_m = 6 \times 10^{-6}$$

$$a_1 = 3.5 \times 10^{-3}, a_2 = .5.$$

An equilibrium solution is found using fictitious play where Japan, South Korea, and China take turns responding optimally to the others, and play stops once no player can benefit from a unilateral move. Each optimal response is performed by maximizing the player's utility function (4) subject to the subgame constraints (5b)-(5e). Optimal MCS provision was proven to be 0 so can be ignored (Theorem 3), leaving fishing quotas, $F_k$, as the only decision variable. When the player owns only one fishery the utility function is quadratic and can be optimized using quadratic binary programming software. The equilibrium solution is displayed in Table 1 along with the corresponding biomasses and utilities. For comparison, the "legal solution" each player would employ to maximize utility if illegal encroachments were not an option is also displayed.

Table 1
**Example 1 Quotas, Biomass, and Utilities**

|       | Legal  | Equilibrium |
|-------|--------|-------------|
| $F_J$ | 792    | 989         |
| $F_S$ | 833    | 1185        |
| $F_C$ | 900    | 1596        |
| $x_J$ | 1.2083 | 0.7453      |
| $x_S$ | 1.1667 | 0.752       |
| $x_C$ | 1.1    | 0.732       |
| $u_J$ | 0.376  | 0.1951      |
| $u_S$ | 0.4167 | 0.2977      |
| $u_C$ | 0.486  | 0.5094      |

Each player is fishing substantially more than the legal optimum. The intuitive explanation for this is that if one player, say South Korea, were to use the legal optimum of $F_S = 833$, then South Korean biomass becomes relatively high and Japan and/or China can issue excessive quotas to induce their fishermen to enter South Korean waters; while this depletes their own biomass and thus utility earned from domestic fishing, the illegal encroachments create a net benefit. In response, South Korea issues excessive quotas to make its waters less desirable to illegal fishermen; this allows them to reap harvests which, even if (legally) suboptimal, would otherwise be extracted by illegal fishermen. This concept is illustrated graphically in Figure 1, where as South Korean quotas are reduced from the equilibrium value to the legal optimum, Chinese quotas increase, and consequently total fishing in South Korean waters decreases by less than the decrease in South Korean quotas (Japanese quotas are held fixed at the equilibrium value). In particular, when $F_S$ is decreased from the equilibrium value of 1185 to the legal optimum of 833 (a 352 quota difference), total fishing in South Korean waters declines by only 75.

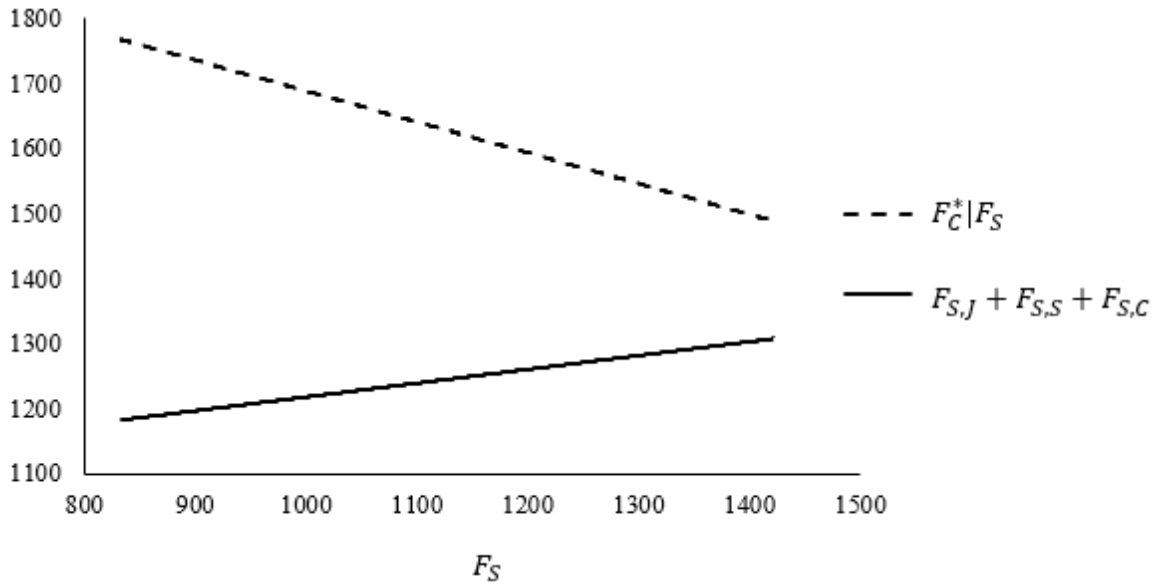

**Fig. 1. Total fishing in South Korean waters as a function of South Korean quotas ($F_S$).**
Note. Japanese quotas are held fixed at $F_J = 989$. As $F_S$ is decreased by 352 from 1185 to 833, total fishing in South Korean waters decreases by only 75.

Another key takeaway from this example is that China's utility at equilibrium is substantially higher than the legal optimum (.5094 vice .486, a 4.81% increase), while Japanese and South Korean utilities are substantially reduced (by 48.11% and 28.56%, respectively). The disparity comes from the fact that while all players are issuing excessive quotas and diminishing their domestic harvests, only Chinese fishermen are fishing illegally in foreign waters, thereby recouping lost domestic harvests. This is seen in Table 2 where the subgame equilibrium (values of $F_{i,k}$) is displayed. Off-diagonals indicate illegal fishing. Only Chinese fisherman fish illegally because they face lower opportunity costs and have more well-equipped maritime militias ($\beta_C < \beta_J, \beta_S$), enabling them to compete in foreign waters despite the additional costs imposed by patrols.

Table 2
**Example 1 Subgame Equilibrium**

| Nationality | Fishery | | |
|---|---|---|---|
| | Japan | South Korea | China |
| Japan | 989 | 0 | 0 |
| South Korea | 0 | 1185 | 0 |
| China | 265 | 63 | 1268 |

The sum of utilities earned by the players at equilibrium is only 89.61% of that under the legal optimum, indicating a potential for mutual gain through bargaining. To assess the benefits of cooperation the following Nash bargaining problem is solved:

$$\max_{F_{i,k},\ m_k} (u_J - u_J^{NC})^{\alpha_J}(u_S - u_S^{NC})^{\alpha_S}(u_C - u_C^{NC})^{\alpha_C} \tag{6}$$

$$s.t.\ u_J \geq u_J^{NC},\ u_S \geq u_S^{NC},\ \text{and}\ u_C \geq u_C^{NC}.$$

The variables $u_k^{NC}$ are the "threat values" for the players, defined as the utilities earned under the noncooperative equilibrium. $\alpha_k \in [0,1]$ represent the bargaining power of players and satisfy $\alpha_J + \alpha_S + \alpha_C = 1$. The decision variables, $F_{i,k}$ and $m_k$ are defined as before, with the critical distinction that the fishing values $F_{i,k}$ now represent legally defined quotas, even when $i \neq k$. For instance, as part of the negotiations Japan may allow a certain number of Chinese fishermen to legally fish in its waters.

Four instantiations of the parameters $\alpha_k$ were used and the results are presented in Table 3. The optimization (6) was estimated numerically using the nested partitioning method (Shi and Ólafsson 2000). Utilities behave in the expected ways, where all players earn greater than the threat values in Table 1 (by necessity), and utilities increase with bargaining power. Interestingly, even in the case where Japan has very strong bargaining power ($\alpha_J = .6$), they cannot attain utility equal to their legal optimum of $.376$. This is because Japan's threat value from which negotiations start, $.1951$, is so far beneath the legal optimum. South Korea likewise cannot attain its legal optimum. China, of course, far exceeds the its legal optimum, as China was already earning more than legality under the noncooperative equilibrium. These dynamics may seem unfair, and perhaps politically unamenable, thus hindering real-world negotiations. Instead of accepting the terms suggested by the Nash bargaining problem, Japan or South Korea may alternatively invest in additional patrols to make Chinese illegal fishing less profitable, improving their utilities under either equilibrium of a cooperative agreement. The viability of this strategy is assessed in the next subsection where comparative statics are performed. To preserve space in Table 3 the solutions are not fully displayed, though the dollar amounts spent on MCS are provided. Note whereas MCS spending was provably 0 in the noncooperative case, the players are now spending a considerable amount on MCS (recall these are measured in billions of USD).

Table 3
Nash Bargaining Results

| Description | $\alpha_J$ | $\alpha_S$ | $\alpha_C$ | $u_J$ | $u_S$ | $u_C$ | $a_1 m_J^{a_2}$ | $a_1 m_S^{a_2}$ | $a_1 m_C^{a_2}$ |
|---|---|---|---|---|---|---|---|---|---|
| *Equal bargaining powers* | 0.33 | 0.33 | 0.33 | 0.2629 | 0.3727 | 0.5982 | 0.0243 | 0.0205 | 0.0212 |
| *Strong China* | 0.25 | 0.25 | 0.5 | 0.2484 | 0.3542 | 0.6459 | 0.0214 | 0.0250 | 0.0185 |
| *Strong Japan* | 0.5 | 0.25 | 0.25 | 0.3006 | 0.3563 | 0.5777 | 0.0188 | 0.0250 | 0.0204 |
| *Strong Japan, weak South Korea* | 0.6 | 0.1 | 0.3 | 0.3213 | 0.3201 | 0.5893 | 0.0258 | 0.0206 | 0.0191 |

*4.2. Comparative statics for Example 1*

Among the more influential parameters in the model are the baseline fishing cost for of the players. In particular, by virtue of China's low costs, $c_C$, China is able to issue excessive quotas to implicitly encourage illegal fishing and achieve utility in excess of its legal optimum. To analyze the sensitivity to $c_C$, Figure 2.a plots the legal and noncooperative equilibrium for each player as it's increased from .0001 to .0002. Unsurprisingly, as China's costs rise its utility declines while Japan's and South Korea's increase. It's interesting to note China only earns more than the legal optimum when $c_C < .000136842$; this does not mean illegal encroachments stop beyond this point. For instance, when $c_C = .00017$, at equilibrium China is still fishing in each Japanese and South Korean waters at levels of $F_{J,C} = 164$ and $F_{S,C} = 19$. China's equilibrium utility falling below the legal optimum merely means the benefits of illegal encroachments are not outweighing the costs of depleted domestic biomass caused by excessive quotas. China nevertheless purposefully depletes its own biomass, as if they do not then the other nations' fishermen will encroach on their waters.

Section 4.1 alluded to the possibility of investment in additional patrols as a means of negating China's illegal encroachments and asymmetric advantage in negotiations. Figure 2.b plots utilities as Japan's investment in patrols, $P_J$, ranges from 10 to 100 patrols boats per 100,000 square kilometers. For reference, a comprehensive empirical study on the use of patrol craft found the most any country employs per 100,000 square kilometers is 100 (Petrossian 2015). While there are gains to be had from increased patrols, the results aren't overly sensitive to this parameter. If Japan were to increase its investment from $P_J = 20$ to $P_J = 100$, equilibrium utility would increase by 18.22%. A more moderate increase to $P_J = 30$ results in a 5.50% gain in utility. Determining whether these investments are worthwhile is challenging for at least two reasons. First, investment in patrols yields benefits outside of their effect on illegal fishing; patrols are valuable in search and rescue missions, port security, and combating trafficking in persons, to name a few areas. Second, rather than investing in more of the existing patrols, a country can invest in force multiplying technologies such as aircraft and surveillance systems that make patrols more effective; this may be a more cost-effective alternative, and may have ancillary benefits unrelated to patrol effectiveness.

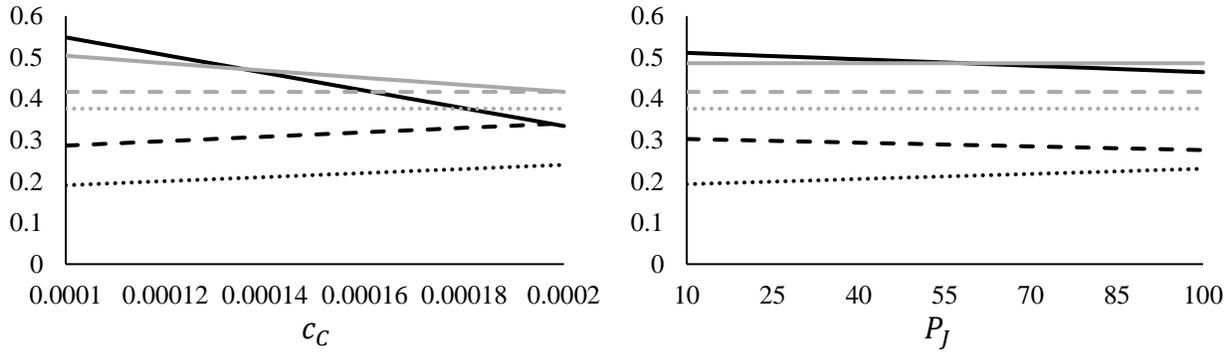

Fig. 2. Comparative statics in Example 1.

Fig. 2.a. Sensitivity to China's baseline costs

Fig. 2.b. Sensitivity to Japanese patrols.

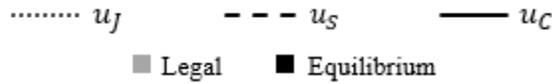

*4.3. Example 2: the legal optimum as an equilibrium*

Example 1 and the corresponding comparative statics showed examples where players fish excessively to either benefit from, or mitigate the damage from, illegal encroachments. It's interesting to consider under what conditions the equilibrium solution may instead correspond to the legal optimal solution. Conditions for this to occur when there are three players, each owning one fishery, are stated in Theorem 4, and an example follows.

**Theorem 4. Conditions for the legal optimum to be an equilibrium.**

Assume a three-player game where each player owns one fishery. Assume $x_k^{OA} < x_k^{m*}$ $\forall k, m \in \{J, S, C\}$ (notation is defined within the proof in Appendix A). Each player using the legally optimal level of fishing quotas represents an equilibrium to the congested environment fishing game if and only if the following conditions hold (all notation is defined in the proof in Appendix A):

**Condition 1.** $p_k q_k x_k^{legal} - c_k > p_m q_m x_m^{legal} - c_k - \beta_k P_m \; \forall \; k, m \in \{J, S, C\}$, where $k \neq m$.

**Condition 2.** For all permutations of $k, l, m$ taking on the values in $\{J, S, C\}$,

$$u_k^{legal} > \max_{F_k \in [F_k^l, F_k^{k,m}]} u_k(F_k) \text{ if } x_k^l > x_k^m. \text{ In addition, } u_k^{legal} > \max_{F_k > F_k^{k,m}} u_k(F_k) \text{ if }$$

$$x_k^m > x_k^{l,m}, \text{ while } u_k^{legal} > \max_{F_k > F_k^{l,m}} u_k(F_k) \text{ if } x_k^m < x_k^{l,m}.$$

*Proof.* See Appendix A.

An example where the legal solution is an equilibrium occurs when the same parameter values as Example 1 are used, with the following exceptions:

$Z_J = Z_C = 1.5, Z_S = 1.25$

$c_J = 210 \times 10^{-6}, c_S = 200 \times 10^{-6}, c_C = 190 \times 10^{-6}$

$P_J = P_S = P_C = 100$

$\beta_J = 18 \times 10^{-7}, \beta_S = 18 \times 10^{-7}, \beta_C = 12 \times 10^{-7}$.

The lower carrying capacities, $Z$, make fishing less lucrative and thus discourage illegal encroachments. Costs being closer to parity removes the comparative advantage China had, which had been contributing to their illegal encroachments. The increased patrols obviously discourage encroachments, as do the larger values of $\beta_k$. To illustrate the dynamics of this example, Figure 3 plots China's utility and the values of subgame variables $F_{JC}$, $F_{SC}$, and $F_{CC}$ as $F_C$ is increased, while fixing $F_J$ and $F_S$ at Japan's and South Korea's legal optima. As China increases its quotas, utility increases up to the legal optimum (attained at $F_C = 789$) before beginning to decline. Eventually, Chinese utility again begins to increase very modestly when $F_C = 1034$, as Chinese biomass is depleted and fishermen begin encroaching on Japanese waters. However, it resumes declining before reaching the legal maximum. This same pattern repeats when $F_C = 1392$ and Chinese fishermen begin encroaching on South Korean waters.

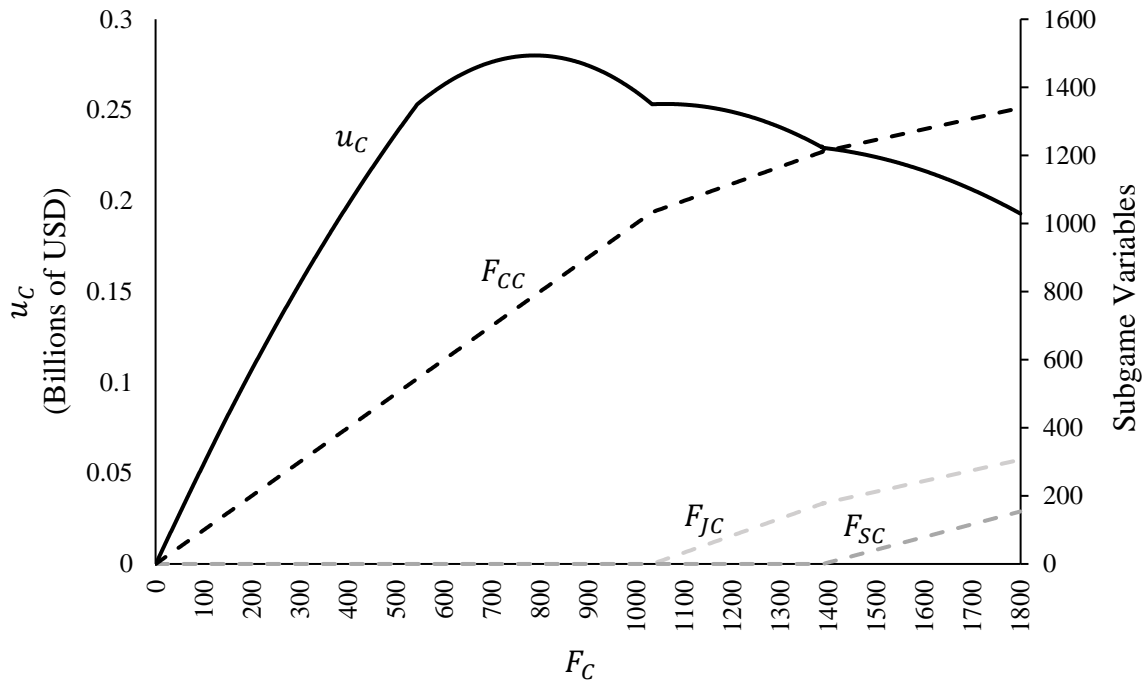

**Fig. 3. Chinese utility and subgame variables for Example 2.**
Note. Japanese and South Korean quotas are held fixed at the legal optima of $F_J = 767$ and $F_S = 733$.

*4.4. Example 3: two fisheries per player*

This final example presents a case where each country owns two fisheries. The parameter values are identical to those used in Example 1, with the exception that a second fishery is now included for each player which has a lower-valued species: $p_{J2} = p_{S2} = p_{C2} = 1.5$. The parameters $Z, r,$ and $q$ for this second fishery are identical to the first for each player. An equilibrium is again found via fictitious play, and it's seen that the concepts illustrated in Example 1 all carry over into this multi-fishery example. Namely: each player issues quotas in excess of the legal optimum; MCS is not used at equilibrium (as proven by Theorem 3); and China achieves utility in excess of its legal optimum on account of illegal encroachments. Table 4 provides the legal and equilibrium solutions. One distinctive feature of this example is that positive MCS is now a part of the legal optimum. When foreign encroachments are not a possibility, MCS serves no purpose when a state owns only one fishery; in contrast, with multiple fisheries a state still uses MCS (and patrols) to ensure its fishermen remain in the fishery for which they've been issues a quota.

**Table 4**
**Example 3 Solutions and Utilities**

|  | Legal | Equilibrium |
|---|---|---|
| $F_{J1}$ | 814 | 641 |
| $F_{J2}$ | 564 | 770 |
| $F_{S1}$ | 855 | 1388 |
| $F_{S2}$ | 646 | 800 |
| $F_{C1}$ | 919 | 1376 |
| $F_{C2}$ | 782 | 1854 |
| $a_1 m_J^{a_2}$ | 0.0227 | 0 |
| $a_1 m_S^{a_2}$ | 0.0224 | 0 |
| $a_1 m_C^{a_2}$ | 0.0224 | 0 |
| $u_J$ | 0.455 | 0.1323 |
| $u_S$ | 0.5272 | 0.3234 |
| $u_C$ | 0.6553 | 0.7246 |

Table 5 shows the legal, equilibrium, and Nash bargaining utilities achieved for additional instantiations of the parameters. These examples use different values of $Z_k$, $P_k$, and $c_C$, and use $\alpha_J = \alpha_S = \alpha_C = 1/3$ for all Nash bargaining problems, (6). The results are not surprising, with utilities increasing in $Z_k$, and improving for Japan and South Korea (and declining for China) as the disparity in patrols and costs declines. One interesting result which was not seen in previous examples is that it's possible for one player, namely Japan, to receive 0 utility under the noncooperative equilibrium. This occurs when carrying capacities are low, and China has a strong asymmetric advantage in either opportunity costs or patrols (see the first three rows of Table 5). In these cases, relatively little rent is available to begin with, and China's asymmetric cost advantages are sufficient to eliminate all Japanese profit potential.

Table 5
Additional Examples with Two Fisheries per Player

| Z | $c_C$ | P | $u_J^{legal}$ | $u_S^{legal}$ | $u_C^{legal}$ | $u_J^{NC}$ | $u_S^{NC}$ | $u_C^{NC}$ | $u_J^{NBS}$ | $u_S^{NBS}$ | $u_C^{NBS}$ |
|---|---|---|---|---|---|---|---|---|---|---|---|
| I | I | I | 0.1021 | 0.1345 | 0.2307 | 0.0000 | 0.0429 | 0.4280 | 0.0280 | 0.0758 | 0.4796 |
| I | I | II | 0.1020 | 0.1376 | 0.2336 | 0.0000 | 0.0929 | 0.3733 | 0.0224 | 0.1227 | 0.4225 |
| I | II | I | 0.1021 | 0.1345 | 0.2307 | 0.0000 | 0.1126 | 0.2063 | 0.0256 | 0.1477 | 0.2426 |
| I | II | II | 0.1020 | 0.1376 | 0.2336 | 0.0456 | 0.1154 | 0.1683 | 0.0653 | 0.1397 | 0.1977 |
| II | I | I | 0.4550 | 0.5272 | 0.6553 | 0.1323 | 0.3234 | 0.7246 | 0.2775 | 0.4930 | 0.9382 |
| II | I | II | 0.4572 | 0.5291 | 0.6571 | 0.1597 | 0.3126 | 0.6800 | 0.3243 | 0.4965 | 0.8947 |
| II | II | I | 0.4550 | 0.5272 | 0.6552 | 0.1740 | 0.3603 | 0.4829 | 0.3470 | 0.5400 | 0.6744 |
| II | II | II | 0.4572 | 0.5291 | 0.6571 | 0.3379 | 0.4235 | 0.4958 | 0.4282 | 0.5117 | 0.6038 |
| III | I | I | 0.8735 | 0.9549 | 1.0937 | 0.3665 | 0.5897 | 1.0358 | 0.6596 | 0.8798 | 1.4151 |
| III | I | II | 0.8752 | 0.9564 | 1.0951 | 0.3924 | 0.5840 | 0.9752 | 0.7252 | 0.9303 | 1.3100 |
| III | II | I | 0.8735 | 0.9550 | 1.0937 | 0.4289 | 0.6501 | 0.7868 | 0.7260 | 0.9470 | 1.1454 |
| III | II | II | 0.8751 | 0.9564 | 1.0951 | 0.5828 | 0.7425 | 0.8759 | 0.7717 | 0.9596 | 1.0844 |

Note. $u_k^{legal}$ represents the legally optimal utilities, $u_k^{NC}$ to those earned at the noncooperative equilibrium, and $u_k^{NBS}$ to those earned from the Nash bargaining problem (6). The columns $Z$, $c_C$, and $P$ indicate different instantiations of model parameters. $Z = $ I corresponds to $Z_{ij} = 1 \; \forall \; i,j$, $Z = $ II to $Z_{ij} = 2 \; \forall \; i,j$, and $Z = $ III to $Z_{ij} = 3 \; \forall \; i,j$. $c_C = $ I corresponds to $c_C = 120 \times 10^{-6}$, while $c_C = $ II corresponds to $c_C = 180 \times 10^{-6}$. $P = $ I corresponds to $[P_J, P_S, P_C] = [20,30,50]$, and $P = $ II corresponds to $[P_J, P_S, P_C] = [100,100,150]$.

## 5. Conclusion and Future Work

This paper modeled a scenario in the ECS, where Japan, South Korea, and China operate fishing fleets in close proximity and face the decisions of how many legal fishing quotas to issue and how much to invest in MCS. Consistent with observed behavior, the model predicts illegal encroachments on account of excessive issuance of quotas. Also consistent with observed behavior, each state underinvests in MCS. The paper thus provides a rational explanation for illegal encroachments and substandard MCS beyond the typical explanations of uncontrollable criminal activity and a lack of political will to implement strong MCS. A bargaining problem was also solved which quantified the substantial gains to be had from cooperation. In the bargain, the player with lower costs, China, is seen to achieve more than the optimal utility they could achieve in the absence of a nearby foreign fisheries, while Japan and South Korea achieve less than this legal optimum. China's gain is realized through legal quotas in the others' waters.

The Gordon-Schaeffer fisheries model with logistic growth and a constant-effort harvesting rate was used throughout this paper. Future research ought to perform similar analysis with alternative fisheries models. Additional complicating factors should also to be assessed, such as migration of species, discounted payoffs, and stochastic growth and harvest rates. Problems with many more than two fisheries per player, and perhaps more than three players, are perhaps the most critical point of future research as such scenario are required to truly model the ECS and other congested maritime environments (the SCS, for instance, is a congested environment with no less than six nations). These many-fishery scenarios are necessary to make specific policy recommendations, and their analysis will require sophisticated computational techniques.

A few other lines of future work are pertinent. As mentioned in Section 2, an empirical analysis of the effectiveness of maritime militias is lacking in the literature. Such a study would improve the analysis presented here by giving evidence-based assessments of $\beta_k$. Accounting for illegal third-parties, such as long-distance fishermen entering the ECS and unloading their catch at far-off ports would also add realism to the model. Lastly, congested maritime environments where there is not a legal delineation of fishing rights is an important area of study. Such is the situation in the SCS, for example, where the consequences of opposing nations' patrol craft confronting one another becomes an important consideration that wasn't relevant in this paper. Not only do interacting patrols likely reduce the deterrent effect of patrols (as fishermen can sound a distress call if confronted by opposition patrols), but the threat of escalation to greater conflict may increase.

**Appendix A. Proofs of Theorems 1 through 4.**

*A.1. Proof of Theorem 1*

**Theorem 1. Existence of a subgame equilibrium.**

Any instantiation of the game's parameters and choice of the overall game's decision variables yields a subgame equilibrium.

*Proof.*

Consider the following algorithm which takes the parameters, $F_{kl}$, and $m_k$ as given, and seeks values of the SGVs:

**Algorithm A1. Finding a SGE.**

1. Initiate all SGVs to $F_{ij,kl} = 0$.

2. Determine the set of SGVs, $F'$, to increase in the next iteration of the algorithm as follows. Define the maximum achievable rent as:

$\pi' = \max\{0, \max\{\pi_{ij,kl} \mid \sum_{ij} F_{ij,kl} < F_{kl}\}\}$; if no $kl$ exists satisfying $\sum_{ij} F_{ij,kl} < F_{kl}$, then set $\pi' = 0$.

If $\pi' = 0$, stop. Otherwise, include all SGVs in $F'$ whose corresponding rent equals $\pi'$, and can be increased without decreasing another SGV; that is, $F_{ij,kl}$ may only be in $F'$ if $\sum_{ij} F_{ij,kl} < F_{kl}$.

Before describing step 3, note the following properties are maintained throughout the algorithm:

**Algorithm A1, Property 1.** SGVs are never allowed to go below 0, and $\sum_{ij} F_{ij,kl} \leq F_{kl}$ remains true for all $kl$.

**Algorithm A1, Property 2.** If $F_{ij,kl} > 0$, then $\pi_{ij,kl} \geq 0$ and $\pi_{ij,kl} \geq \pi_{ij',kl}$ for all $ij'$.

These two properties ensure all subgame constraints are satisfied throughout the algorithm, other than those stating no profitable quotas are left unused. The latter condition is the stopping condition for the algorithm ($\pi' = 0$).

3. Increase all SGVs in $F'$ while obeying the following rules:

   i. Increase SGVs in $F'$ at rates such that their rents remain equal.

   ii. Consider the case where $\sum_{ij} F_{ij,kl} = F_{kl}$ for some $kl$ (so that by definition $F_{ij,kl}$ are not in $F'$). Modify the SGVs $F_{ij,kl}$ at rates such that $\sum_{ij} F_{ij,kl} = F_{kl}$ remains true, and $F_{ij,kl} > 0$ iff $\pi_{ij,kl} \geq \pi_{ij',kl}$ for all $ij'$.

   iii. Return to step 2 whenever a new SGV enters $F'$, or when $\sum_{ik} F_{ij,kl'} = F_{kl'}$ for any SGVs in $F'$, or when the rent of a SGV in $F'$ becomes 0.

The proof this algorithm always produces a SGE is clear, after noting one additional property to accompany Properties 1 and 2.

**Algorithm A1, Property 3.** $\pi'$ is either constant or decreasing linearly throughout the algorithm. The former case occurs only when SGVs in $F'$ are increasing with a perfectly offsetting decrease in SGVs not in $F'$, a process which eventually stops when one of the decreasing SGV reaches 0, and is followed by a state where the only SGVs being modified are increasing. Thus, every state where $\pi'$ is constant is followed by a state where it's linearly decreasing, and thus $\pi'$ will eventually reach 0.

$\pi' = 0$ implies either all quotas are exhausted, or no fisherman type $kl$ can achieve positive rent with their remaining quotas. Thus, properties 1 through 3 imply the algorithm will terminate at a subgame equilibrium.

*A.2. Proof of Theorem 2*

**Theorem 2. Utilities are uniquely defined by decision variables.**

Any instantiation of the game's parameters and choice of the overall game's decision variables leads to uniquely defined utilities for the players. This is in spite of the fact that multiple subgame equilibria (SGE) may exist.

*Proof.*

This theorem is proved in two parts. First, it's shown that any SGE leads to the same set of biomasses for the fisheries in the game. Next, it becomes easy to show all players receive the same utility under any SGE, given biomasses are the same.

To show any SGE leads to the same biomasses, assume two distinct SGE exist. These will be referred to as SGE1 and SGE2, and the "prime" and "double prime" notion will be used to distinguish their characteristics. That is, $x'_{ij}$ denotes the biomass of fishery $ij$ under SGE1, while $x''_{ij}$ denotes that for SGE2. Assume $x'_{ij} < x''_{ij}$ for some fishery $ij$. A consequence of this is that $F'_{ij,kl} > F''_{ij,kl}$ for some $kl$; otherwise, $x'_{ij} < x''_{ij}$ would be impossible. A further consequence is that fisherman type $kl$ is using all quotas under SGE2: $\sum_{ij} F''_{ij,kl} = F''_{kl}$. This is the case because fishing yields nonnegative rent for $kl$ at biomass level $x'_{ij}$ (otherwise $F'_{ij,kl} > F''_{ij,kl} \geq 0$ would violate the condition that fishermen only fish if profitable), and thus biomass level $x''_{ij} > x'_{ij}$ yields strictly positive rent. Were $kl$ to have available quotas ($\sum_{ij} F''_{ij,kl} < F''_{kl}$), they would be used to extract this positive rent in fishery $ij$, eventually exhausting all quotas.

It can also be shown that every fishery where fisherman type $kl$ is operating under SGE2 has less biomass under SGE1. To see this, denote the set of fisheries where $kl$ was using positive quotas under SGE2 as $S''_{kl} := \{mn \mid F''_{mn,kl} > 0\}$; the claim is that $x'_{mn} < x''_{mn} \; \forall \; mn \in S''_{kl}$. To prove this claim, note the subgame conditions require each fisherman type, $kl$, to earn equal profits everywhere they fish, and that it's known $F'_{ij,kl} > 0$. Now assume $kl$ is fishing in $ij$ under SGE2: $ij \in S''_{kl}$. This immediately reveals the profits $kl$ earns under SGE1 are less than under SGE2 (since $x'_{ij} < x''_{ij}$). If profits are less under SGE1, then all fisheries in $S''_{kl}$ must have less biomass under SGE1; if a fishery existed violating this condition, then fisherman type $kl$ could earn greater profits there and would divert from $ij$ to said fishery. Assume instead $kl$ isn't fishing in $ij$ under SGE2: $ij \notin S''_{kl}$. This simply means that under SGE2, $kl$ had more attractive profits outside of fishery $ij$. These are now gone, since $F'_{ij,kl} > 0$ despite $x'_{ij,kl} < x''_{ij,kl}$; these more attractive profits being "gone" is equivalent to stating biomasses have declined. In sum, it's now been shown all fisheries in $S''_{kl}$ must have experienced a decline in biomass.

A recursive argument extending the above analysis will lead to a contradicting, thus showing $x'_{ij} < x''_{ij}$ is impossible. First note that there must be some fishery $i_2 j_2 \in S''_{kl}$ such that another fisherman type, say $k_2 l_2$, is fishing more under SGE1 than under SGE2: $F'_{i_2 j_2, k_2 l_2} > F''_{i_2 j_2, k_2 l_2}$ for some $i_2 j_2 \in S''_{kl}$. This is the case because fisherman type $kl$ alone cannot cause a decline in every fishery in $S''_{kl}$, as if it were possible to do this profitably then SGE2 wouldn't be a subgame equilibrium. Now repeating the previous logic, because $k_2 l_2$ is fishing (more) in $i_2 j_2$ under SGE1 and $x'_{i_2 j_2} < x''_{i_2 j_2}$, all fisheries where $k_2 l_2$ is fishing under SGE2 (denote these $S''_{k_2 l_2}$) have lower biomass under SGE1. Also by the previous logic: (i) $k_2 l_2$ is using all quotas under SGE2; and (ii) fisherman type $k_2 l_2$ can't cause the lower biomasses in $S''_{k_2 l_2}$ alone. More importantly for the recursive argument, however, is that the lower biomasses for each fishery in the union $S''_{kl} \cup S''_{k_2 l_2}$ cannot be caused by increased levels of fishing from $kl$ and $k_2 l_2$ alone. This is because each is using all available quotas under SGE2. A third fisherman type, $k_3 l_3$, must be expending more quotas in at least one fishery from $S''_{kl} \cup S''_{k_2 l_2}$ under SGE1 than under SGE2. Repeating the previous line of reasoning leads to a recursion ending in a contradiction: $k_3 l_3$ is using all quotas under SGE2, all fisheries in $S''_{k_3 l_3}$ have experienced a decline in biomass, and fisherman types $kl$, $k_2 l_2$, and $k_3 l_3$ alone can't cause the biomass declines in $S''_{kl} \cup S''_{k_2 l_2} \cup S''_{k_3 l_3}$; an additional fisherman type must be contributing to the declines; repeating, eventually there are no additional fisherman types to explain biomass declines, and thus a contradiction has been found. The conclusion is the original assumption, that $x'_{ij} < x''_{ij}$ for some $ij$, is an impossibility. Thus, all SGE have the same biomasses for all fisheries.

Having established that all SGE lead to identical biomasses, it's easy to show all SGE lead to the same utilities for all players. Recall that all fisherman types, $kl$, receive equal profits in all fisheries where they're fishing. Assume type $kl$ isn't using all available quotas under a particular SGE, which means the best available profit at the biomass levels for this SGE is less than or equal to 0. In this case, fisherman type $kl$ contributes 0 to the utility of Player $k$, regardless of exactly where they are fishing (i.e. regardless of the specific values of $F_{ij,kl}$ under this SGE).

Assume instead that $kl$ is using all available quotas. A condition of the subgame is that fishermen fish only where profits are highest. Denote the maximal profit achievable by fisherman type $kl$ by the constant $v$, and note this is the same across all SGE since biomasses are the same for all SGE. Thus, under any SGE the contribution of fisherman type $kl$ to Player $k$'s utility is $F_{kl} \cdot a$, a constant. It's thus been shown all SGE yield identical utilities for the players.

*A.3. Proof of Theorem 3*

**Theorem 3. Non-use of MCS.**

Assume each player owns one fishery. Consider Player $k$'s decision. For given strategies of all other player, responding with $m_k > 0$ can yield at most equivalent utility as $m_k = 0$. If costs of MCS are nonzero, then $m_k > 0$ yields strictly less utility than $m_k = 0$.

*Proof.*

Because each player owns only one fishery, denote the number of fishermen from country $k$ fishing in country $i$'s waters as $F_{i,k}$. Assume player $k$ is using non-zero MCS: $m_k > 0$. Unilaterally changing strategy to $m_k = 0$ only affects the costs imposed on fishermen from country $k$ fishing in foreign waters. For all $i \neq k$, if $F_{i,k}$ doesn't change at subgame equilibrium on account of the change in $m_k$ (because it was not previously profitable to illegally fish in $i$, and is still not), then all SGVs are unchanged and $k$'s utility is either the same (if MCS is costless) or has increased (if MCS costs money). If $F_{i,k}$ does change, then it must have increased due to the reduction in costs. This means either: (i) Player $k$ extracts rent from $i$'s waters, improving her utility; or (ii) Player $k$'s fishermen have diverted from $k$'s waters to $i$'s. In the latter case, $k$ can simply issue more quotas to replace those who diverted to $i$'s waters, establishing a subgame equilibrium equivalent to that when $m_k > 0$ but with a higher value of $F_{i,k}$. By the previous logic, Player $k$'s utility has increased. This completes the proof.

*A.4. Proof of Theorem 4*

**Theorem 4. Conditions for the legal optimum to be an equilibrium.**

Assume a three-player game where each player owns one fishery. Assume $x_k^{OA} < x_k^{m^*}$ $\forall\ k, m \in \{J, S, C\}$ (notation is defined within the proof). Each player using the legally optimal level of fishing quotas (i.e. that which would be used if illegal encroachments were not a possibility) represents an equilibrium to the congested environment fishing game if and only if the following conditions 1 and 2 hold (all notation is defined in the accompanying proof):

**Condition 1.** $p_k q_k x_k^{legal} - c_k > p_m q_m x_m^{legal} - c_k - \beta_k P_m$ $\forall\ k, m \in \{J, S, C\}$, where $k \neq m$.

**Condition 2.** For all permutations of $k, l, m$ taking on the values in $\{J, S, C\}$, $u_k^{legal} > \max_{F_k \in [F_k^l, F_k^{k,m}]} u_k(F_k)$ if $x_k^l > x_k^m$. In addition, $u_k^{legal} > \max_{F_k > F_k^{k,m}} u_k(F_k)$ if $x_k^m > x_k^{l,m}$, while $u_k^{legal} > \max_{F_k > F_k^{l,m}} u_k(F_k)$ if $x_k^m < x_k^{l,m}$.

*Proof.*

While the notation used to define the necessary and sufficient conditions is quite cumbersome, the proof straightforward. First note that the legally optimal number of quotas Player $k$ should issue is $F_k^{legal} = \frac{r_k}{2 q_k}\left(1 - \frac{c_k}{p_k q_k Z_k}\right)$. This is easily derived using calculus, noting Player $k$'s biomass is $x_k = Z_k\left(1 - \frac{q_k}{r_k} F_k\right)$ and utility is hence $u_k = (p_k q_k x_k - c_k) \cdot F_k$. Defining $x_k^{legal} =$ $Z_k \left(1 - \frac{q_k}{r_k} F_k^{legal}\right)$, condition 1 simply states that when each player issues the legally optimal quotas, no illegal encroachments occur. Were this condition violated, say because $p_C q_C x_C^{legal} - c_C < p_J q_J x_J^{legal} - c_C - \beta_C P_J$, then Chinese fishermen would have an incentive to fish illegally in Japanese waters. In turn, China could issue additional quotas to farm Japanese biomass down until $p_C q_C x_C^{legal} - c_C = p_J q_J x_J^{legal} - c_C - \beta_C P_J$, at which point China is reaping the legal harvest from its own waters while reaping additional harvests from Japanese waters, thus obviously achieving utility in excess of the legal optimum.

Condition 1 ensures the only possible way a player could benefit from a unilateral deviation from the legal solution is to farm its own biomass below the legally optimal level, for the purpose of extracting illegal rent from another's waters. Condition 2 ensures this is not a viable strategy, as shown next. Without loss of generality, assume $F_J = F_J^{legal}$ and $F_S = F_S^{legal}$, and analyze China's utility, $u_C(F_C)$, as $F_C$ is increased. Denote the level of Chinese biomass such that it becomes economical for Chinese fishermen to divert into the waters of Player $i$ as $x_C^i$. This is easily calculated by solving: $p_C q_C x_C^i - c_C = p_i q_i x_i^{legal} - c_C - \beta_C P_i$. The level of Chinese fishing attaining this biomass, $F_C^i$, can also be easily calculated by solving $Z_C \left(1 - \frac{q_C}{r_C} F_C^i\right) = x_C^i$.

Again without loss of generality, assume $F_C^J < F_C^S$, so that as $F_C$ is increased Chinese fishermen encroach on Japanese waters before South Korean. Once $F_C > F_C^J$, each additional unit of Chinese quotas increases actual fishing in Chinese waters, $F_{CC}$, by less than 1. Denote the increase from a unit increase in $F_C$ as $\Delta F_{CC}'$, which can be easily computed by solving a linear system of 2 equations and 2 unknowns: $\Delta F'_{CC} + \Delta F'_{JC} = 1$; $\frac{p_C q_C^2}{r_C} \Delta F'_{CC} = \frac{p_J q_J^2}{r_J} \Delta F'_{JC}$. The latter equation ensures Chinese fishermen continue to receive equal rent in Chinese and Japanese waters. For any level of Chinese quotas beyond $F_C^J$ and up to an upper-bound (to be specified momentarily), Chinese biomass is $x'_C = x_C^J - \frac{z_C q_C}{r_C}(F_C - F_C^J)\Delta F'_{CC}$, and by equality of rent all Chinese fishermen receive $p_C q_C x'_C - c_C$, and hence Chinese utility is simply $u'_C = (p_C q_C x'_C - c_C)F_C$.

The upper-bound on this utility function's applicability is the value of $F_C$ such that Chinese and Japanese biomasses have been so degraded that either Chinese or Japanese fishermen find it profitable to start fishing in South Korean waters. There's no guarantee who will start encroaching on South Korean waters first, so each possibility will be enumerated. It's already been stated that Chinese fishermen will encroach on South Korean waters once $x_C = x_C^S$, and Japanese fishermen will once $x_J = x_J^S$. To frame everything in terms of Chinese biomass, $x_J^S$ can be converted to a corresponding biomass in Chinese waters using the fact Chinese fishermen are receiving equal rents. Denoting $x_C^{J,S}$ as the level of Chinese biomass where Japanese fishermen begin encroaching on South Korean waters, equality of Chinese rents implies: $p_C q_C x_C^{J,S} = p_J q_J x_J^S - \beta_C P_J \to x_C^{J,S} = \frac{p_J q_J x_J^S - \beta_C P_J}{p_C q_C}$. Also denote the corresponding level of Chinese fishing $F_C^{J,S}$, which is found by solving: $F_C^{J,S} = F_C^J + a$, where $a = \frac{(x_C^J - x_C^{J,S})}{\frac{z_C q_C}{r_C}\Delta F_{CC}} = \frac{r_C(x_C^J - x_C^{J,S})}{z_C q_C \Delta F_{CC}}$, the additional level of fishing to deplete Chinese biomass from $x_C^J$ to $x_C^{J,S}$, provided that $x_C^{J,S} > x_C^S$. Likewise, the level of fishing needed to deplete Chinese biomass to $x_C^S$ is $F_C^{C,S} = F_C^J + \frac{r_C(x_C^J - x_C^S)}{z_C q_C \Delta F_{CC}}$. The upper-bound on the applicability of utility function $u'_C$ is the lesser of $F_C^{C,S}$ and $F_C^{J,S}$; optimizing $u'_C$ within the bounds $\left[F_C^J, \min\{F_C^{C,S}, F_C^{J,S}\}\right]$ is a straightforward, one-dimensional quadratic maximization ($u'_C$ is quadratic because $x'_C$ is linear in $F_C$).

Now assume $F_C^{C,S} < F_C^{J,S}$, so that Chinese fishermen begin encroaching on South Korean waters first, and assess China's utility function for $F_C > F_C^{C,S}$. Beyond $F_C^{C,S}$, a unit increase in $F_C$ changes $F_{CC}$ by $\Delta F''_{CC}$, which is computed by solving a linear system of 3 equations and 3 unknowns: $\Delta F''_{CC} + \Delta F''_{JC} + \Delta F''_{SC} = 1$; $\frac{p_C q_C^2 z_C}{r_C} \Delta F''_{CC} = \frac{p_J q_J^2 z_J}{r_J} \Delta F''_{JC}$; and $\frac{p_C q_C^2 z_C}{r_C} \Delta F''_{CC} = \frac{p_S q_S^2 z_S}{r_S} \Delta F''_{SC}$. The latter two constraints ensure Chinese fishermen continue to receive equal rent in all fisheries. Chinese biomass for $F_C > F_C^{C,S}$ is therefore $x''_C = x_C^S - \frac{z_C q_C}{r_C}(F_C - F_C^{C,S})\Delta F''_{CC}$, and utility is $u''_C = (p_C q_C x''_C - c_C) F_C$, which is again quadratic in $F_C$. Unlike $u'_C$, the domain of $u''_C$ has no upper-bound on $F_C$, provided China does not find it optimal to use so many quotas that either Japan or South Korea can no longer earn any profits in its own waters. This is because as Chinese quotas increases, by necessity Chinese fishermen's rents decrease by equal amount, hence revenues ($p_i q_i x_i$), which don't depend on who is doing the fishing, decrease by equal amounts, and hence neither Japanese nor South Korean fishermen will change behavior for some critical value of $F_C$. To formalize the assumption that Japan and South Korea continue to earn domestic profits as China seeks its optimal response to $F_J^{legal}$ and $F_S^{legal}$, it's assumed $x_k^{OA} < x_k^{C*}$ for $k \in \{J, S\}$, where $x_k^{C*}$ is the level of biomass corresponding to China's optimal response given $F_C > F_C^{C,S}$ and $x_k^{OA}$ is the "open access" level of Player $k$'s biomass, where their rents are fully depleted: $p_k q_k x_k^{OA} = c_k \to x_k^{OA} = \frac{c_k}{p_k q_k}$. Finding China's optimal utility subject to $F_C > F_C^{C,S}$ is again a straightforward quadratic maximization.

Lastly, assume $F_C^{C,S} > F_C^{J,S}$ so Japanese fishermen begin encroaching on South Korean waters first. For this case, when $F_C > F_C^{J,S}$ the change in Chinese fishing in Chinese waters, $\Delta F_{CC}'''$, caused by a unit increase in $F_C$ can be found by solving a system of four linear equations and four unknowns: $\Delta F_{CC}''' + \Delta F_{JC}''' = 1; \frac{p_C q_C^2 z_C}{r_C} \Delta F_{CC}''' = \frac{p_J q_J^2 z_J}{r_J}(\Delta F_{JC}''' + \Delta F_{JJ}'''); \frac{p_S q_S^2 z_S}{r_S} \Delta F_{SJ}''' = \frac{p_J q_J^2 z_J}{r_J}(\Delta F_{JC}''' + \Delta F_{JJ}'''); \Delta F_{SJ}''' = -\Delta F_{JJ}'''$. The second and third constraints ensure Chinese and Japanese fishermen, respectively, maintain equality of rent in the waters they fish in, and the last constraint reflects the fact the total quotas issued by Japan has not changed. Similarly to previous cases, Chinese biomass can be written as $x_C''' = x_C^{J,S} - \frac{z_C q_C}{r_C}(F_C - F_C^{J,S})\Delta F_{CC}'''$, and China's utility function for $F_C > F_C^{J,S}$ is $u_C''' = (p_C q_C x_C''' - c_C)F_C$, which is quadratic and easily maximized.

To summarize the proof, when $F_J = F_J^{legal}$ and $F_S = F_S^{legal}$, China cannot improve upon $F_C = F_C^{legal}$ from a unilateral change, assuming $F_C^J < F_C^S$, if:

**Condition 1**

- $p_C q_C x_C^{legal} - c_C > p_m q_m x_m^{legal} - c_C - \beta_C P_m$ for $m \in \{J, S\}$.

**Condition 2**

- $\max_{F_C \in [F_C^J, \min\{F_C^{C,S}, F_C^{J,S}\}]} u_C' < u_C^{legal}$,

- $\max\limits_{F_C > F_C^{C,S}} u_C'' < u_C^{legal}$ if $F_C^{C,S} < F_C^{J,S}$, and

- $\max\limits_{F_C > F_C^{J,S}} u_C''' < u_C^{legal}$ if $F_C^{C,S} > F_C^{J,S}$.

Analogous statements can be made for the case where $F_C^J > F_C^S$, as well as for Japan and South Korea seeking improvements on their legally optimal levels of quotas. This completes the proof.

**Appendix B. Comments on parameter selection.**

The primarily purpose of this paper was to present an explanation for observed behavior in the ECS, backed by analytical results that apply to any congested maritime environment. While detailed parameter estimation of specific fisheries was not conducted, parameters were chosen to be sensible, as described in this appendix.

*B.1. Operating and opportunity costs, $c_k$*

Operating costs will vary wildly by vessel type, but this paper assumed a value of 50,000 USD, informed by Cabral et al. (2018) which includes cost estimates of vessels of approximately 1000 gross tons (GT). Assuming 10 fishermen per vessel, an assumption on annual salaries provides an estimate of total operating plus opportunity costs. Salaries for fishermen by country were taken from https://www.salaryexpert.com/salary/browse/countries/fisherman-deep-sea. Chinese annual salaries were 14,700 USD, South Korean 34,752 USD, and Japanese 57,624 USD. Final figures were rounded off while maintaining orders of magnitudes.

*B.2. Price of fish, $p_{ij}$*

As with fishing costs, prices vary significantly (this time by species). Tai et al. (2017) estimated the average price of fish used for direct human consumption to be 1,750 USD per ton in 2010. Considering price increases from inflation and the general depletion of fish stocks, and the variation in high-quality and low-quality species, the parameter values of 3,000 USD and 1,500 USD per ton used in this paper are reasonable.

*B.3. Biological parameters $Z_{ij}$, $q_{ij}$, and $r_{ij}$*

These parameters again vary significantly, based on the specific fishery being modeled. This paper didn't model specific fisheries (e.g. tuna, herring, etc.), so did not have much to base parameter values on. They were chosen to be relatively aligned to those found in Chen and Andrew (1998). To generalize results, multiple examples were presented with differing values of $Z_{ij}$. Examples for alternative $r_{ij}$ values weren't presented because the results would have been similar to shocking $Z_{ij}$; a slower natural growth rate has a similar affect to a smaller carrying capacity in the Gordon-Schaefer model. Alternative values of $q_{ij}$ weren't presented because the fundamental insights would continue to hold. A higher $q_{ij}$ implies fewer fishermen are needed to reap the same harvest; however, China would still possess the same asymmetric cost advantages over South Korea and Japan, and would use this to issue excessive quotas which encroach on foreign fisheries. In turn, South Korea and Japan also issue excessive quotas rather than allowing China to extract profits unimpeded, as before.

*B.4. Maritime law enforcement parameters $P_k$, $β_k$, $β_m$, $a_1$, and $a_2$*

The choices of patrols per 100,000 square kilometers of EEZ, $P_k$, were informed by (Petrossian 2015) and Petrossian (2019). Only Japan is listed explicitly with a value of $P_J = 20$, which is close to the average value of 20.28. It was assumed South Korea would be slightly above average (hence $P_S = 30$ was used in Example 1), and that China would be significantly above ($P_C = 50$) based on its well-document patrol investments in recent years (Erickson 2018). When patrols were shocked to higher values in Figure 2.b and Table 5, the maximum value found in Petrossian of 100 (Taiwan) was kept in mind.

$β_k$ and $β_m$ were also informed by Petrossian (2015), where a linear regression model was used to show each patrols and MCS as significantly explainers of illegal fishing. While linear regression coefficients don't translate directly to the model used in this paper, $β_k$ values were chosen so that costs imposed by patrols amounted to roughly 10-20% of operating and opportunity costs. MCS was shown to be a more decisive deterrent in Petrossian (2015), and thus $β_m$ was chosen such that no Chinese fishermen would enter South Korean waters illegally in Example 1 when $m_S = 100$ and $x_S = 1.1667$ (the legal optimum in Example 1). That is, the following was solved:

$p_S q_S x_S = c_C + β_C P_S + 100 β_m \rightarrow β_m = 6 \times 10^{-6}$ (after rounding to the nearest significant digit).

Lastly, the shape parameter for the cost of MCS, $a_2$, was arbitrarily set to .5. The parameter $a_1$ was then selected so that the cost of $m_S = 100$ was similar to the cost spent by Norway (Mangin et al. 2018), whom the FAO has rated highly for quality of MCS and which has similar sized fisheries to those used in this paper (Pitcher, Kalikoski, and Pramod 2006). This led to $a_1 100^{a^2} = .034 \rightarrow a_1 = 3.5 \times 10^{-3}$.